\documentclass[sigconf]{acmart}
\usepackage{bm}
\usepackage{amsmath}
\usepackage{booktabs}
\usepackage{multirow}
\usepackage{multicol}
\usepackage[normalem]{ulem}
\useunder{\uline}{\ul}{}
\usepackage{appendix}
\usepackage{balance}
\usepackage{makecell}
\usepackage{color}
\usepackage[marginal]{footmisc}
\usepackage[linesnumbered,ruled,vlined, noend]{algorithm2e}
\usepackage{subfigure}

\usepackage{caption}
\usepackage{graphicx}
\usepackage{float} 
\usepackage{algorithmic}
\usepackage{marvosym}

\usepackage{enumitem}
\setitemize[1]{topsep=0pt}

\newcommand{\stitle}[1]{\vspace{1ex} \noindent{\bf #1}}

\AtBeginDocument{%
  \providecommand\BibTeX{{%
    \normalfont B\kern-0.5em{\scshape i\kern-0.25em b}\kern-0.8em\TeX}}}

\begin{document}

% \title{Poisoning Attacks against Contrastive Recommender Systems}
% \title{Threats Hidden in Contrastive Recommender Systems: \\Poisoning Attack Amplifies Representation Dispersion}
\title{Unveiling Vulnerabilities of Contrastive Recommender Systems to Poisoning Attacks}

\author{Zongwei Wang}
\email{zongwei@cqu.edu.cn}
\orcid{0000-0002-9774-4596}
\affiliation{%
  \institution{Chongqing University}
  \country{China}
}

\author{Junliang Yu}
\email{jl.yu@uq.edu.au}
\affiliation{%
  \institution{The University of Queensland}
  \country{Australia}}
  
\author{Min Gao}
\email{gaomin@cqu.edu.cn}
\authornote{Corresponding author}
%\thanks{*Corresponding author}
\affiliation{%
  \institution{Chongqing University}
%%  \streetaddress{1 Th{\o}rv{\"a}ld Circle}
  \country{China}}

\author{Hongzhi Yin}
\email{h.yin1@uq.edu.au}
\authornotemark[1]
\affiliation{%
  \institution{The University of Queensland}
  \country{Australia}}

\author{Bin Cui}
\email{bin.cui@pku.edu.cn}
\affiliation{%
  \institution{Peking University}
  \country{China}}

\author{Shazia Sadiq}
\email{shazia@eecs.uq.edu.au}
\affiliation{%
  \institution{The University of Queensland}
  \country{Australia}}

% \affiliation{%

% \affiliation{%

% \affiliation{%

% \affiliation{%

% \affiliation{%

% \affiliation{%

% \renewcommand{\shortauthors}{Z and Tobin, et al.}
\renewcommand{\shortauthors}{Zongwei Wang et al.}

\begin{abstract}
Contrastive learning (CL) has recently gained prominence in the domain of recommender systems due to its great ability to enhance recommendation accuracy and improve model robustness. Despite its advantages, this paper identifies a vulnerability of CL-based recommender systems that they are more susceptible to poisoning attacks aiming to promote individual items. Our analysis indicates that this vulnerability is attributed to the uniform spread of representations caused by the InfoNCE loss. Furthermore, theoretical and empirical evidence shows that optimizing this loss favors smooth spectral values of representations. This finding suggests that attackers could facilitate this optimization process of CL by encouraging a more uniform distribution of spectral values, thereby enhancing the degree of representation dispersion. With these insights, we attempt to reveal a potential poisoning attack against CL-based recommender systems, which encompasses a dual-objective framework: one that induces a smoother spectral value distribution to amplify the InfoNCE loss's inherent dispersion effect, named dispersion promotion; and the other that directly elevates the visibility of target items, named rank promotion. We validate the threats of our attack model through extensive experimentation on four datasets. By shedding light on these vulnerabilities, our goal is to advance the development of more robust CL-based recommender systems. The code is available at \url{https://github.com/CoderWZW/ARLib}.
\end{abstract}

\ccsdesc[500]{Information systems~Recommender systems}

\keywords{Contrastive Learning, Recommender Systems, Poisoning Attacks, Self-Supervised Learning}

\copyrightyear{2024}
\acmYear{2024}
\setcopyright{acmlicensed}
\acmConference[KDD '24]{Proceedings of the 30th ACM SIGKDD Conference on Knowledge Discovery and Data Mining}{August 25--29, 2024}{Barcelona, Spain}
\acmBooktitle{Proceedings of the 30th ACM SIGKDD Conference on Knowledge Discovery and Data Mining (KDD '24), August 25--29, 2024, Barcelona, Spain}
\acmPrice{15.00}
\acmDOI{10.1145/3580305.3599324}
\acmISBN{979-8-4007-0103-0/23/08}


\maketitle

\section{Introduction}
In recent years, contrastive learning (CL) \cite{27jaiswal2020survey, 28khosla2020supervised} has emerged as a promising self-supervised learning paradigm in deep representation learning, demonstrating its significant potential across various domains \cite{35you2020graph, 36shuai2022review}. This approach encourages models to learn informative features and invariance from vast amounts of unlabeled data, achieving enhancement in the model's generalization ability and leading to improved accuracy and robustness in different tasks. When applied to recommender systems, CL has been extensively studied and proven effective in enhancing recommendation quality \cite{06yu2023self,25yu2022xsimgcl}. Additionally, its utilization in this domain exhibits a remarkable level of robustness comparable to that observed in computer vision \cite{02wang2022robustness,11ghosh2021contrastive} and natural language processing \cite{01fan2021does,12choi2022c2l}, confirming its ability to uphold recommendation accuracy even in the presence of noisy data scenarios \cite{08wu2021self,03ye2023towards,71wang2023efficient}.

\begin{figure*}[t]
\centering
\includegraphics[width=0.95\textwidth]{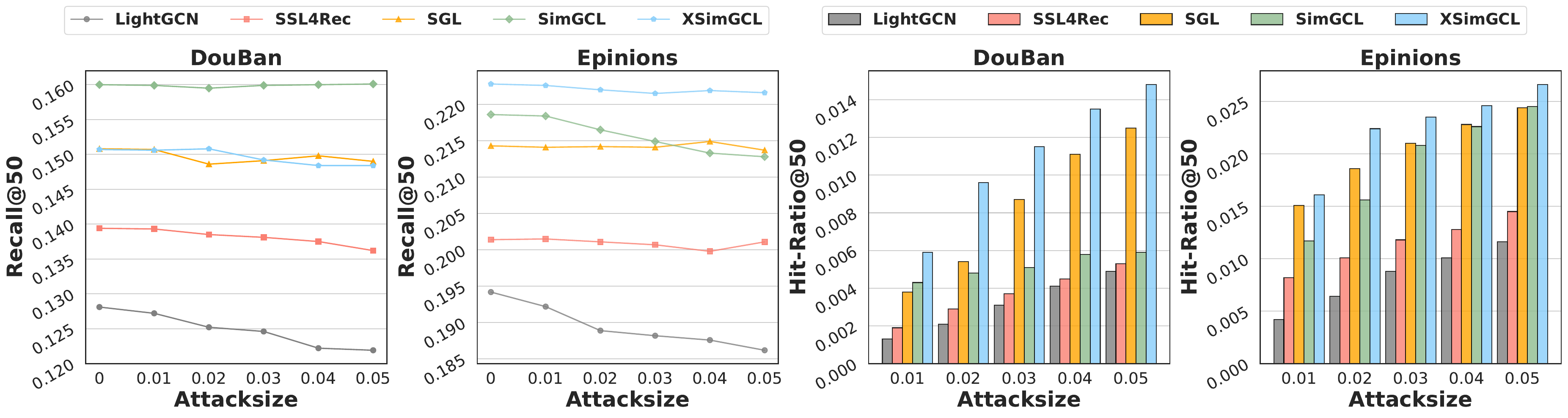}
% \vspace{-5pt}
\caption{The comparison of LightGCN, SSL4Rec, SGL, SimGCL, and XSimGCL on DouBan and Epinions under Random Attack. Attack size represents the ratio of the number of malicious users to the total number of users.} 
\label{randomattack}
\end{figure*}

However, it is noteworthy that existing studies predominantly concentrate on the overall robustness of CL-based recommendation, overlooking the vulnerabilities affecting individual items. For example, in the context of product recommendations, there have been cases where certain low-quality and unpopular products are intentionally promoted to gain profits \cite{04wang2022gray}. Similarly, in the domain of news recommendations, misleading information is disseminated to specific groups on purpose \cite{30yi2022ua}. It is conceivable that the successful promotion of target items with ulterior motives can have a significantly negative impact on user experience and even social stability. Within the realm of recommender systems, one prominent type of attack, referred to \textit{poisoning attacks}, can focus on individual items  \cite{13zhang2022pipattack, 29li2016data,68zheng2024poisoning,69yuan2023manipulating,70yin2024device}. This type of attack fulfills deliberate manipulation of recommendation outcomes through strategic injection of malicious profiles, whose ultimate goal is to amplify the exposure of target items. As poisoning attacks are widespread and pose a genuine threat in many real-world recommender systems, a natural question arises: \textit{Can CL-based recommendation effectively withstand the poisoning attacks on individual target items?}

To address our concerns, we first conducted a series of experiments specifically focused on poisoning attacks against recommender systems, comparing the performance of recommendation models with and without CL. For our experiments, we employed the widely used model, LightGCN \cite{07he2020lightgcn}, as the recommendation encoder, and evaluated four popular and easily reproducible CL-based recommendation methods: SSL4Rec \cite{55yao2021self}, SGL \cite{08wu2021self}, SimGCL \cite{09yu2022graph} and XSimGCL \cite{25yu2022xsimgcl}, which differ in the way of constructing contrastive views. To initially explore the impact of poisoning attacks, we adopted a common attack approach (Random-Attack) \cite{10lam2004shilling} to inject randomly constructed malicious user profiles into public datasets. Regrettably, as illustrated in Figure~\ref{randomattack}, despite the ability of CL-based recommendation methods to maintain stability in overall recommendation accuracy (measured by Recall@50), the results revealed that they were all more susceptible to poisoning attacks that aim to increase the exposure of target individual items (measured by Hit-Ratio@50). This unexpected outcome has sparked a profound sense of worry and urgency within us, as it implies that the widespread adoption of CL-based recommendation may have caught the considerable attention of potential attackers.

To identify the underlying causes of CL-based recommendation's vulnerability against poisoning attacks, we investigate the embedding space learned by non-CL and CL-based recommendation methods. By visualizing the distributions of the representations, we find that the objective function of CL, InfoNCE \cite{37oord2018representation}, is the core factor. In the absence of CL, the representations show \textit{local clustering} characteristics, where users and popular items tend to gather and hinder cold items from reaching more users. In contrast, the addition of CL provides repulsion between any two nodes, thus suppressing the aggregation of users and popular items and leading to \textit{globally dispersed} representation distribution. Our findings reveal that CL is a double-edged sword. On the one hand, it has an inhibitory effect on popular items and then alleviates the popularity bias problem \cite{26abdollahpouri2019popularity}. On the other hand, non-popular items could be easier to enter recommendation lists. As target items in poisoning attacks are often unpopular, CL becomes a booster to increase the exposure of target items. This inherent flaw gives rise to a new question: \textit{Does there exist a potential poisoning attack that poses a greater threat to CL-based recommendation than current poisoning attacks?}

In this paper, we give an affirmative answer to this question. Given that the CL objective serves as an auxiliary task relative to the primary recommendation objective, the representation distribution from the joint optimization (i.e., the attraction endowed by the recommendation objective and the repulsion induced by the CL objective) might not be sufficiently uniform to meet the attacker's criteria. Potential adversaries may thus choose to amplify the dispersion degree of representation distribution, ensuring users are more broadly dispersed across the embedding space. This increased dispersion creates a conducive attack environment, providing target items with expanded opportunities to reach more diverse users.

Based on the above insights, we theoretically and empirically establish that the fine-tuning of CL correlates with smooth spectra of representations. This revelation supports the notion that attackers can facilitate the optimization of CL loss by promoting a smoother spectral value distribution, consequently increasing the degree of dispersion. On this basis, an attack model is proposed to uncover the poisoning mechanism against CL-based recommendation, aiming to deepen the understanding of the robustness of CL-based recommender systems and safeguard them from manipulation. Specifically, we explore a bi-level optimization based attack framework against CL-based recommendation, named the \underline{C}ontrastive \underline{Lea}rning \underline{R}ecommendation Attack (CLeaR). The CLeaR pursues two primary objectives: Firstly, the dispersion promotion objective focuses on achieving a smoother spectral distribution of representations to facilitate optimizing CL loss, thereby enhancing the extent of their dispersion across the vector space. Secondly, the rank promotion objective is dedicated to optimizing the visibility of target items, ensuring their maximized exposure to an extensive user base.

Our contributions can be summarized as follows:
\begin{itemize}[leftmargin=*]
\item We delve into the vulnerability of CL-based recommender systems to poisoning attacks, which has still remained unexplored despite the popularity of CL-based recommendation.
\item We propose a potential poisoning attack strategy to examine the detrimental impact of poisoning attacks on CL-based recommendation.
\item Extensive experiments on several datasets illustrate that the proposed attack method carries a greater threat than existing attack methods to both general and CL-based recommendation. 
\end{itemize}

\begin{figure*}[t]
  \centering
  \includegraphics[width=1\linewidth]{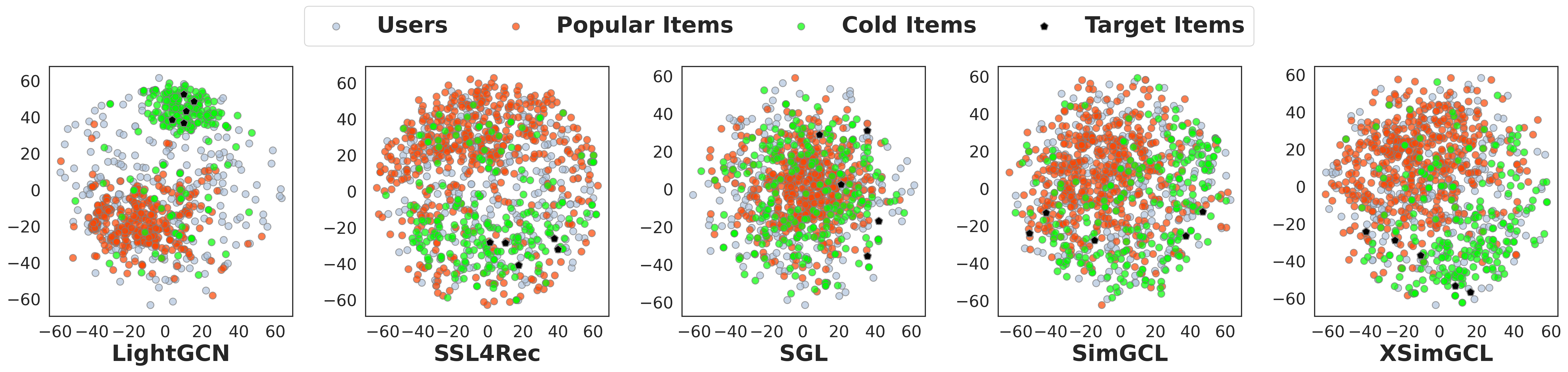} 
  \caption{Representation distribution on Epinions under poisoning attacks (Random-Attack). Recommendation methods without CL show a local clustering pattern, while the ones with CL show a global dispersion pattern.}
\label{Epinions_embeding}
\end{figure*}

\section{PRELIMINARIES}
\subsection{Recommendation Models}
\stitle{Recommendation Task.}
Let $\mathcal{U} $ and $\mathcal{I} $ respectively be the sets of users and items, and $\mathcal{D}$ denote the user interaction data. A recommendation model aims to learn a low-dimensional representation set $\mathbf{Z}$ for predicting the preference of user $u \in \mathcal{U} $ to item $i \in \mathcal{I}$, which can be calculated by $\mathbf{z}_{u}$ and $\mathbf{z}_{i} \in \mathbf{Z}$. A widely used recommendation objective function is Bayesian Personalized Ranking (BPR) \cite{15rendle2012bpr}, which is formulated as:  
\begin{equation}
  \mathcal{L}_{rec}= \underset{(u, i, j) \sim P_{\mathcal{D}}} {\mathbb{E}} -\log \, \phi (\mathbf{z}_u^T \mathbf{z}_i- \mathbf{z}_u^T \mathbf{z}_j),
  \label{BPRloss}
\end{equation}
where $\phi$ is the sigmoid function. $P_{\mathcal{D}}(\cdot)$ refers to the distribution of the interaction data, the tuple $(u, i, j)$ includes a user $u$, a positive sample item $i$ with observed interactions, and a negative sample item $j$ without observed interactions with $u$. 

\stitle{Contrastive Learning Task.}
CL \cite{16chen2020simple} aims to maximize the consistency between augmentations of the original representations (e.g., learned from perturbed data). The most commonly used contrastive
loss is InfoNCE \cite{37oord2018representation} formulated as follows:
\begin{equation}
\begin{aligned}
\mathcal{L}_{cl} = -\sum_{p \in (\mathcal{U} \cup \mathcal{I})} \log \frac{\exp\left( {\overline{\textbf{z}}'_p}^T \overline{\textbf{z}}''_p / \tau \right)}{\sum_{n \in \mathcal{(\mathcal{U} \cup \mathcal{I})}} \exp \left( {\overline{\textbf{z}}'_p}^T \overline{\textbf{z}}''_{n}/ \tau \right)},
\end{aligned}
\label{CLloss}
\end{equation}
where $\overline{\textbf{z}}_{n}$ represents the negative sample from $\mathcal{U}$ or $\mathcal{I}$, $\overline{\textbf{z}}$ is the $\text{L}_{2}$ normalized representations of $\textbf{z}$, $\overline{\textbf{z}}'_{p}$ and $ \overline{\textbf{z}}''_{p}$ are learned from different augmentations of the entity $p$, and $\tau $ is the hyper-parameter, known as the temperature coefficient. 

 % if $n \neq p$ and $n$ and $p$ belong to the same mini-batch $\mathcal{B}$
Generally, CL-based recommendation models optimize a joint loss where the recommendation is the primary task while the CL
task plays a secondary role whose magnitude is controlled by a
coefficient $\omega$. The joint loss is formulated as follows:
\begin{equation}
\mathcal{L}= \mathcal{L}_{rec}+ \omega \mathcal{L}_{cl}.
\end{equation}

\subsection{Poisoning Attacks}
\stitle{Attacker's Goal.} 
In recommender systems, poisoning attacks are of two kinds: non-targeted attacks aiming to degrade overall performance, and targeted attacks intended to promote specific items. Our attack falls under the category of targeted promotion, where we aim to maximize the likelihood of target items $\mathcal{I}^{T}$ appearing in as many users' top-K recommendation lists as possible.

\stitle{Attacker's Prior Knowledge.} There are two cases for attackers to know about victim models. In the white-box setting, attackers have full access to the recommendation model. While in the black-box setting, attackers only have access to interaction information $\mathcal{D}$ and do not possess any prior knowledge about the recommendation model. To figure out the attack mechanism against CL-based recommender systems and investigate the theoretical upper bound of potential damage, our attack strategy is primarily concentrated on the white-box setting. %It is noteworthy that the adaptation of a white-box attack strategy to a black-box environment often entails the use of a proxy model. This approach has been extensively acknowledged and validated in current research \cite{45wu2021ready, 46zhang2020practical}.

\stitle{Attacker’s Capability.} Attackers can inject a set of malicious users $\mathcal{U}_{M}$ and generate well-designed interactions $\mathcal{D}_{M}$. However, due to limited resources, the attacker can only register a restricted number of malicious users with limited interactions. If not stressed, we assume that the attacker can register malicious users equivalent to 1\% of the total number of genuine users, and the interaction count of each malicious user can not be more than the average number of interactions per normal user. This setting is commonly used in related studies \cite{05gunes2014shilling,04wang2022gray,13zhang2022pipattack}.

\section{How Poisoning Attacks Affect CL-Based Recommendation}
Given the fact that CL-based recommender systems are more vulnerable to poisoning attacks aimed at target item promotion, this section will explore the cause behind the susceptibility.

\subsection{Global Dispersion Affects Local Clustering}

Drawing inspiration from prior research \cite{09yu2022graph, 18wang2021understanding, 19wang2020understanding}, we first visualize the representation distributions learned by recommendation models with and without CL when subjected to a poisoning attack. The visualization results on Epinions dataset, as shown in Figure \ref{Epinions_embeding}, reveal two distinct patterns in the absence and presence of CL:  
\begin{itemize}[leftmargin=12pt]
    \item Local Clustering (without CL): recommendation without CL (LightGCN) exhibits a tendency to cluster popular items and users together, while cold items are positioned distantly from the user clusters.
    \item Global Dispersion (with CL): in contrast, recommendation with CL (SSL4Rec, SGL, SimGCL, and XSimGCL) exhibits a more balanced distribution, extending throughout the entire vector space. Consequently, this distribution brings popular items and cold items into closer proximity.
\end{itemize}

The details on the design of visualization experiments, along with additional visualization results on the DouBan \cite{32Zhao} dataset are available in Appendix \ref{Visualization Experiment}. 
Furthermore, the emergence of two distinct patterns (i.e., local clustering and global dispersion) from empirical observations can be explained from the loss optimization perspective.

\stitle{BPR Loss Causes Local Clustering.}
The local clustering pattern has been discovered in many existing studies \cite{09yu2022graph, 13zhang2022pipattack}. The gradient update formulas of BPR loss from Equation~\eqref{BPRloss} are as follows:
\begin{equation}
  \nabla_{\mathbf{z}_u} = - (\mathbf{z}_i - \mathbf{z}_j) \cdot \frac{\exp(\mathbf{z}_u^T \mathbf{z}_i - \mathbf{z}_u^T \mathbf{z}_j)}{1 + \exp(\mathbf{z}_u^T \mathbf{z}_i - \mathbf{z}_u^T \mathbf{z}_j)}.
  \label{BPRgradient1}
\end{equation}

By examining Equation \eqref{BPRgradient1}, the gradient update direction ($-\nabla_{z_u}$) pulls users towards positive item $i$ while distancing themselves from negative item $j$. Given the long-tail distribution characteristic of recommendation data, popular items have a higher likelihood of being sampled. Consequently, users are more inclined to be attracted to popular items, leading to a clustering effect between users and popular items. In contrast, cold items are often treated as negative samples and are pushed away from the user clusters. 

\stitle{CL Loss Causes Global Dispersion.}
Recall Equation~\eqref{CLloss}, we can derive the following format:
\begin{small}
\begin{equation}
\begin{aligned}
\mathcal{L}_{cl} = - (\sum_{p \in \mathcal{(U\cup I)}} (\overline{\mathbf{z}}'^T_p \overline{\mathbf{z}}''_p / \tau)  - \sum_{p \in \mathcal{(U\cup I)}} \log \sum_{n \in \mathcal{(U\cup I)}} \exp (\overline{\mathbf{z}}'^T_p \overline{\mathbf{z}}''_{n}/ \tau)),
\end{aligned}
\label{cllosstransformation}
\end{equation}
\end{small}and we can deduce that minimizing CL loss involves maximizing the similarity between $\mathbf{z}'_{p}$ and $\mathbf{z}''_{p}$, while simultaneously minimizing the similarity between $\mathbf{z}'_{p}$ and $\mathbf{z}''_{n}$. We attribute global dispersion to the second term of the formula: this process of CL loss optimization can be perceived as pushing $\mathbf{z}'_{p}$ and $\mathbf{z}''_{n}$ away from each other. Consequently, the CL loss inherently generates a natural repulsion among the representations, causing the representation to be roughly uniformly distributed.

\stitle{Discussion.} 
The CL loss appears to be a double-edged sword in recommendation. On the one hand, it helps to mitigate the popularity bias problem. The global dispersion pattern caused by CL loss provides cold items with an opportunity to be exposed to a larger user base. This observation is supported by Figure ~\ref{Epinions_embeding} and and ~\ref{embeding}, which show that many cold items are in close proximity to more users in CL-based recommendation methods. On the other hand, the CL loss breaks down the local clustering pattern, inadvertently facilitating the targeting of specific items through poisoning attacks. Particularly, considering that the target items are often cold items, this property of CL gives these items more exposure to a larger number of users through the poisoning attack. This raises a pertinent question of whether attackers can intensify the global dispersion patterns caused by CL loss. In a scenario where such optimization is accentuated, target items would have a higher visibility in the recommendation lists of a wider user base. In the following subsection, we will explicate the correlation between the optimization of CL loss and the spectral characteristics of representations.

\subsection{Analysis of CL Loss from a Spectral Perspective}
In this section, we first theoretically show that the optimization of CL loss favors smooth representations' spectral values (also called singular values). Then, we empirically show the distribution of singular values decomposed by representation learned by recommendation models with and without CL.

\subsubsection{\textbf{Theoretical Analysis.}}
Inspired by \cite{52liu2022revisiting}, we consider the decomposition of representations and use Singular Value Decomposition (SVD) \cite{54hoecker1996svd} to get the decomposition of $\textbf{Z}$, i.e., $\textbf{Z} = \mathbf{L} \mathbf{\Sigma} \mathbf{R}^T$, where $\mathbf{L} = (\textbf{l}_{1}, ..., \textbf{l}_{d})$ represents $d$-order orthogonal matrix and $\mathbf{R} = (\textbf{r}_{1}, ..., \textbf{r}_{N})$ represents $N$-order orthogonal matrix. The terms $\textbf{l}$ and $\textbf{r}$ correspond to left singular vector and right singular vector, respectively. Here, $d$ is the dimensional length of the representation vector, while $N$ is the number of all users and items. $\mathbf{\Sigma} = diag(\mathbf{\sigma}_{1}, ..., \mathbf{\sigma}_{d})$ is the $d \times N$ rectangular diagonal matrix, with non-negative singular values $\sigma$ arranged in descending order. Then, we can get an upper bound of the CL loss:

\stitle{Proposition 1}. \textit{Give the representations $\textbf{Z}^\prime$ \textit{and} $\textbf{Z}^{\prime\prime}$ which are learned on augmented views} \textit{and the corresponding singular values } $\mathbf{\Sigma}' = diag(\mathbf{\sigma}'_{1}, ..., \mathbf{\sigma}'_{d})$ \textit{and} $\mathbf{\Sigma}'' = diag(\mathbf{\sigma}''_{1}, ..., \mathbf{\sigma}''_{d})$, \textit{an upper bound of the CL loss is given by:}
\begin{equation}
\begin{aligned}
\mathcal{L}_{cl} < N \max_{j} \sigma'_{j}  \sigma''_{j} -\sum_{i} \sigma'_{i} \sigma''_{i} + N \log N.
\end{aligned}
\end{equation}

\textit{Proof.} \textit{See Appendix} \ref{Proof of Proposition 1}.

Since the minimizing the CL loss is equivalent to minimizing this upper bound, Proposition 1 suggests that the minimization of the CL loss can be transformed into a dual-objective optimization problem: minimizing the first term $\max_{j} \sigma'_{j}  \sigma''_{j}$, which corresponds to the largest singular value product; and maximizing the second term $\bm{\vec{\sigma}}'^{T}\bm{\vec{\sigma}}''$, which corresponds to the sum of the product of singular values. This approach targets the reduction of the highest singular value similarities while enhancing the overall similarity of the singular values between $\textbf{Z}'$ and $\textbf{Z}''$. Such a minimization strategy is inclined to yield a smoother distribution of singular values. Next, we will empirically show the difference of singular values derived from CL-based and non-CL recommendation models.

\subsubsection{\textbf{Empirical Observations.}} In the experimental setup analogous to prior configurations, representations are derived from LightGCN, SSL4Rec, SGL, SimGCL, and XSimGCL. Subsequent to this, SVD is employed to decompose these representations into singular values, which are then sorted.

\begin{figure}[t]
  \centering
  \includegraphics[width=1\linewidth]{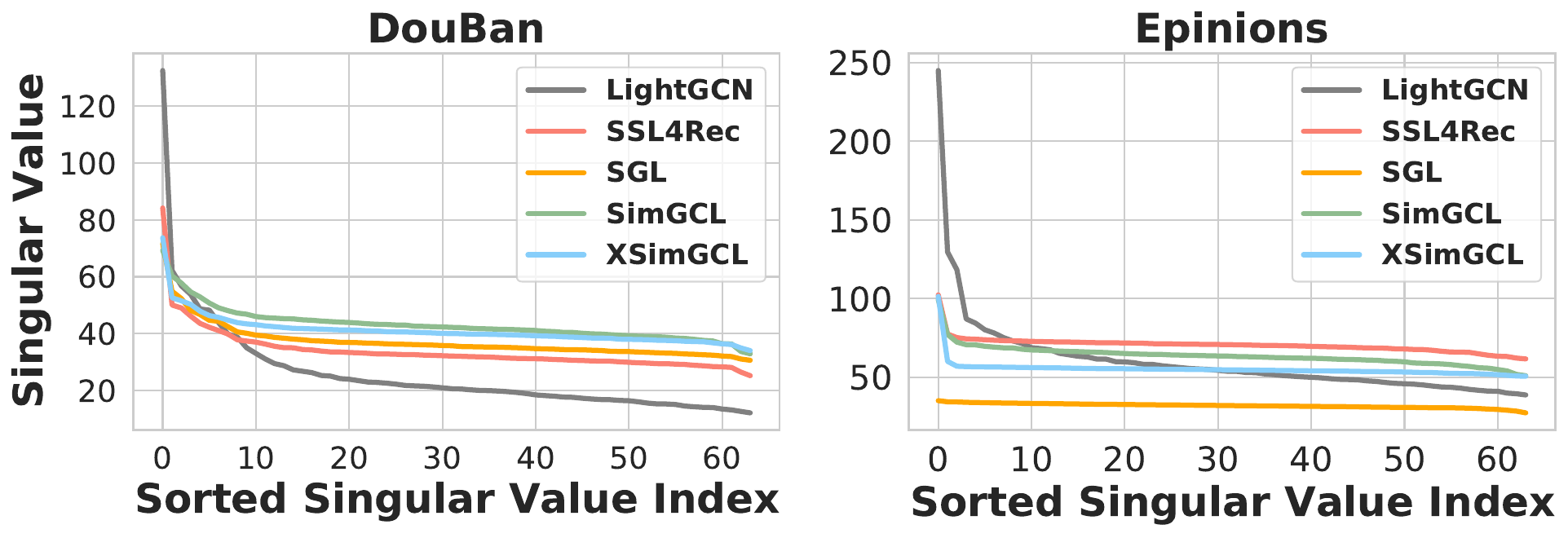}  
  \caption{Singular Value Distributions on two datasets.}
\vspace{-10pt}
\label{spectralvalue}
\end{figure}

Figure \ref{spectralvalue} illustrates the distribution of singular values for both non-CL and CL-based recommendation. Both distributions obey the long-tail distribution. However, notable distinctions can be observed: the LightGCN curves appear sharper across both datasets, whereas the curves corresponding to CL-based recommendation maintain smooth. Our theoretical analysis can explain this phenomenon, i.e., the optimization of CL loss tends to gravitate towards diminishing the maximal spectral value, concurrently facilitating a smoother distribution of spectral values. Therefore, we can observe that the CL-based recommendation models tend to temper larger singular values, simultaneously advocating for a smooth transition between singular values. 

\stitle{Discussion.}
Through empirical and theoretical demonstration, we conclude that the optimization of CL loss can contribute to a smooth spectral value distribution. This view prompts us to wonder whether a poisoning attack can become more powerful by additionally promoting smoother spectral values in representations. Such an approach will favor the optimization of CL loss and in turn disperse user representations. In the next section, we will introduce a novel poisoning attack method that focuses on spectral values of representations to enhance attack effectiveness.

\section{Proposed Poisoning Attacks}
 In this section, we reveal a more powerful poisoning attack, CLeaR, which aims to homogenize spectral values in representations, thereby impacting a broader user base effectively.

\subsection{Attack as a Bi-Level Optimization Problem}
We define the poisoning attack as the bi-level optimization setting: the inner optimization is to derive the recommendation model details based on real user interactions and fixed malicious user interactions set. On the other hand, the outer optimization is to adjust the interactions between malicious users and items, with the goal of increasing the presence of target items in more user recommendation lists. The bi-level optimization process is defined by the following equations: 

\begin{equation}
\begin{aligned}
\max\limits_{\mathcal{D}_{M}}&\mathcal{L}_{attack}(\mathcal{D}, \mathcal{D}_{M}, \mathbf{Z}^{*}_{\mathcal{U}}, \mathbf{Z}^{*}_{\mathcal{I}}), \\
\mathrm{s.t.}, \quad  \mathbf{Z}^{*}_{\mathcal{U}},& \mathbf{Z}^{*}_{\mathcal{I}}=\arg\min\limits_{\mathbf{Z}_{\mathcal{U}}, \mathbf{Z}_{\mathcal{I}}}\mathcal{L}_{rec}(f(\mathcal{D},\mathcal{D}_{M})),
\end{aligned}
\label{bi-level optimization}
\end{equation}
where $f$ is the function of the recommendation model, $\mathbf{Z}^{*}_{\mathcal{U}}, \mathbf{Z}^{*}_{\mathcal{I}}$ are the optimal user representations and item representations, and $\mathcal{L}_{attack}$ is the loss function for evaluating attack utility. It is worth noting that there are no interactions for the initialization of $\mathcal{D}_{M}$, and the interactions are generated as the inner and outer optimizations are alternately performed.

\subsection{CLeaR Overview}

\begin{figure}[t]
\centering
\includegraphics[width=0.4\textwidth]{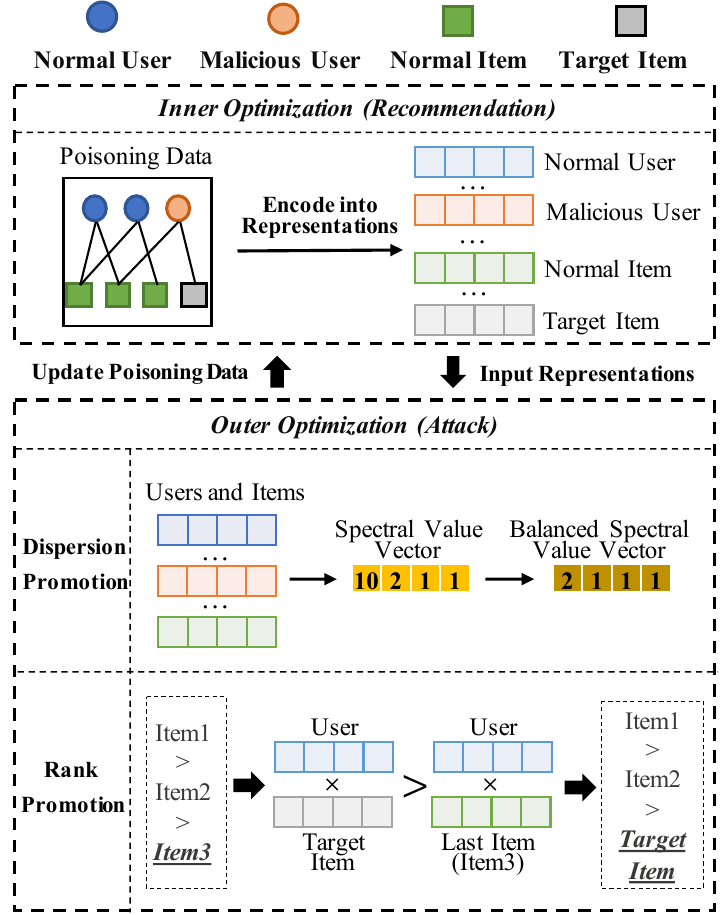}
\caption{The overview of CLeaR.} 
\label{attackmethod}
\label{-10pt}
\end{figure}

To explore the worst-case scenario, we provide the white-box implementation of CLeaR under the bi-level optimization framework. This framework is designed to be adaptable to black-box scenarios through the integration of proxy models. The simple overview can be seen in Figure ~\ref{attackmethod}, encompassing the following steps:

\subsubsection{\textbf{Inner Optimization}} Initially, it is essential to obtain the representations of both users $\mathbf{Z}_{\mathcal{U}}$ and items $\mathbf{Z}_{\mathcal{I}}$, as this serves as the fundamental prior knowledge for subsequent steps. To achieve this, we train the recommendation model to obtain optimal representations as follows:

\begin{equation}
\begin{aligned}
\mathbf{Z}^{*}_{\mathcal{U}}, \mathbf{Z}^{*}_{\mathcal{I}}=\arg\min\limits_{\mathbf{Z}_{\mathcal{U}}, \mathbf{Z}_{\mathcal{I}}}\mathcal{L}(f(\mathcal{D},\mathcal{D}_{M})).
\end{aligned}
\end{equation}

\subsubsection{\textbf{Outer Optimization}} In this step, by establishing interactions between malicious users and target items, as well as between malicious users and the candidate items, the goal of the attack is to promote target items to enter more recommendation lists. To achieve this goal, the CleaR combines two pivotal objective functions: a dispersion promotion objective that indirectly augments the dispersion of representations and a rank promotion objective to pull users into recommendation lists directly.

\stitle{Dispersion Promotion Objective:} This way aims to promote spectral values of representations to be smoother, thereby facilitating the optimization of CL loss and a comprehensive dispersion across the entirety of the vector space. In this case, target items naturally gain augmented opportunities to optimize proximity to broader users. In alignment with this concept, a common way is to first decompose a representation $\textbf{Z}$ to get $\bm{\vec{\sigma}}$ by SVD, then align the spectral values $\bm{\vec{\sigma}}$ of a smooth and flat distribution. Therefore, we initially formulate the dispersion promotion objective $\mathcal{L}_{D}$ as:
\begin{equation}
\begin{aligned}
\mathcal{L}_{D} = sim (\bm{\vec{\sigma}}, cx^{-\beta}),
\end{aligned}
\label{simdistribution}
\end{equation}
where $sim(\cdot)$ calculates the similarity of two distribution. $cx^{-\beta}$ is the power-law distribution, $c$ and $\beta$ are the hyper-parameters. A smaller value of $\beta$ would lead to a gradual transition into the long-tail part of the distribution.

However, the applicability of this method encounters limitations because obtaining the spectrum of a matrix is time-consuming and the backpropagation of SVD is unstable which often results in numerical inaccuracies. To address this challenge, we adopt the a spectral approximation approach, drawing inspiration from rank-1 approximation techniques as discussed in  \cite{56zhang2023spectral,57yu2020toward}, which directly calculates approximated representations $\mathbf{\hat{Z}}$, thereby bypassing the need for the SVD process. The equation of the approximated representation is as follows:
\begin{equation}
\begin{aligned}
\mathbf{\hat{Z}}=\textbf{Z} - \frac{\textbf{Z}\textbf{V}'\textbf{V}'^{T}}{||\textbf{V}'||_{2}^{2}},
\end{aligned}
\end{equation}
where $\textbf{V}'=\textbf{Z}^{T}\textbf{Z}\textbf{V}$, and $\textbf{V}$ is randomly re-initialized in each iteration. This approach can approximately adjust the representation $\textbf{Z}$ by applying a scaling mechanism where larger singular values are diminished using smaller scaling factors, whereas smaller singular values are amplified with larger factors. Consequently, this leads to the approximated representation $\mathbf{\hat{Z}}$ exhibiting a more uniform singular value distribution following its decomposition.

% To solve this problem, we follow the idea from \cite{56zhang2023spectral,57yu2020toward} that aims to target the direct emulation of an approximated balanced spectrum. The approximated balanced spectrum of $\bm{\vec{\sigma}}$ sorted from largest to smallest is as follows:

% \begin{equation}
% \begin{aligned}
% \bm{\vec{\sigma}}'=\bm{\vec{\sigma}}\vec{\gamma},
% \end{aligned}
% \end{equation}
% where $\vec{\gamma} \in (0,1)$ and $\vec{\gamma}$ is the vector with incremental quantity. This operation means that multiplying the term with a larger singular value by a smaller value, and multiplying the term with a smaller singular value by a larger value, thereby getting a flatter vector $\bm{\vec{\sigma}}'$. Then, the equation of the approximated representation is as follows:

% \begin{equation}
% \begin{aligned}
% \textbf{z}'=\textbf{z} - \frac{\textbf{z}\textbf{v}'\textbf{v}'^{T}}{||\textbf{v}'||_{2}^{2}},
% \end{aligned}
% \end{equation}
% where $\textbf{v}'=\textbf{z}^{T}\textbf{z}\textbf{v}$, and $\textbf{v}$ is a random vector.

After getting the $\mathbf{\hat{Z}}$, we can avoid repetitive computations and backpropagation of SVD. The Equation (\ref{simdistribution}) is transformed into calculating a similarity between the original representation $\textbf{Z}$ and the approximated representation $\mathbf{\hat{Z}}$, and in our method, we use $L_{1}$ distance as $sim(\cdot)$. The formulation of the dispersion promotion objective is as follows:
\begin{equation}
\mathcal{L}_{D} = sim (\mathbf{Z}, \mathbf{\hat{Z}}) 
=- \frac{||\textbf{Z}\textbf{V}'\textbf{V}'^{T}||}{||\textbf{V}'||_{2}^{2}}.
\end{equation}

\stitle{Rank Promotion Objective:}
The dispersion promotion objective is strategically crafted to amplify the dispersion of representations. From the perspective of the user, such dispersion merely establishes a level playing field for all items, which reduces the difficulty for target items to be included in the user's recommendation list. Nonetheless, it is important to note that this equalization does not inherently guarantee that the target item will surpass others in terms of recommendation priority. Consequently, there is a need for an objective tailored to boost the ranking of target items in user recommendations, facilitating their presence in the user's list of preferences. Taking inspiration from the latest target attack methods in recommendation, we optimize the differentiable CW loss \cite{20rong2022fedrecattack} as the rank promotion objective. We aim to find the last item in the recommended list and increase the probability of the target items more than such last item. The formulation of the CW loss is: 
\begin{equation}
\begin{aligned}
\mathcal{L}_{R}=&\sum_{u \in (\mathcal{U})}\sum_{t \in \mathcal{I}^{T} \land \left(u, t\right) \notin \mathcal{D}} g(\mathbf{z}_u^T \mathbf{z}_t - \min_{i \in {\mathcal{I}_u^{\text{rec}} \land i\notin \mathcal{I}^{T}}}\left\{\mathbf{z}_u^T \mathbf{z}_i\right\}),\\
&where \quad g(x)= \begin{cases}x, & x \geq 0, \\ e^x-1, & x<0,\end{cases}
\end{aligned}
\end{equation}
where $\min_{i \in {\mathcal{I}_u^{\text{rec}} \wedge i\notin \mathcal{I}^{T}}}\left\{\mathbf{z}_u^T \mathbf{z}_i\right\}$ means the last item in the recommendation list of user $u$, and $\mathcal{I}^{rec}_{u}$ is the set of items recommended for user $u$. Finally, we combine the two promotion objectives, and then the $L_{attack}$ is shown as follows:
\begin{equation}
\begin{aligned}
\mathcal{L}_{attack}=\mathcal{L}_{D}+\alpha \mathcal{L}_{R}.
\end{aligned}
\end{equation}

Based on the aforementioned description, we present the attack algorithm of CLeaR in Appendix \ref{Algorithm}.

\begin{table*}[t]
\caption{Performance comparison of different attack methods on the CL-based RS. The best results are in bold, and the runners-up are with underlines. H and N refer to Hit Ratio and NDCG, respectively. The corresponding values for Yelp should be multiplied by $10^{-3}$.}
\vspace{-10pt}
\centering
\resizebox{\textwidth}{!}{
\begin{tabular}{@{}c|ccc|cc|cc|cc|cc|cc|cc|cc|cc@{}}
\toprule
\multirow{2}{*}{Dataset} &
  \multicolumn{1}{c|}{\multirow{2}{*}{Model}} &
  \multicolumn{2}{c|}{\textbf{RandomAttack}} &
  \multicolumn{2}{c|}{\textbf{BandwagonAttack}} &
  \multicolumn{2}{c|}{\textbf{AUSH}} &
  \multicolumn{2}{c|}{\textbf{DLAttack}} &
  \multicolumn{2}{c|}{\textbf{PoisonRec}} &
  \multicolumn{2}{c|}{\textbf{FedRecAttack}} &
  \multicolumn{2}{c|}{\textbf{A\_hum}} &
  \multicolumn{2}{c|}{\textbf{GSPAttack}} &
  \multicolumn{2}{c}{\textbf{CLeaR}} \\ \cmidrule(l){3-20} 
 &
  \multicolumn{1}{c|}{} &
  H@50 &
  N@50 &
  H@50 &
  N@50 &
  H@50 &
  N@50 &
  H@50 &
  N@50 &
  H@50 &
  N@50 &
  H@50 &
  N@50 &
  H@50 &
  N@50 &
  H@50 &
  N@50 &
  H@50 &
  N@50 \\ \midrule
\multirow{5}{*}{DouBan} &
  \multicolumn{1}{c|}{LightGCN} &
  0.0010 &
  0.0037 &
  0.0011 &
  0.0100 &
  0.0013 &
  {\textbf{0.0106}} &
  0.0012 &
  {\ul 0.0100} &
  {\ul 0.0016} &
  0.0083 &
  0.0013 &
  0.0033 &
  0.0014 &
  0.0026 &
  0.0014 &
  0.0025 &
  \textbf{0.0023} &
  0.0039 \\
 &
  \multicolumn{1}{c|}{SSL4Rec} &
  0.0019 &
  0.0082 &
  0.0034 &
  0.0095 &
  0.0058 &
  0.0094 &
  0.0138 &
  0.0094 &
  0.0146 &
  0.0088 &
  0.0176 &
  0.0095 &
  {\ul 0.0177} &
  {\ul 0.0108} &
  0.0163 &
  0.0100 &
  \textbf{0.0181} &
  \textbf{0.0129} \\
 &
  \multicolumn{1}{c|}{SGL} &
  0.0038 &
  0.0108 &
  0.0020 &
  0.0109 &
  0.0018 &
  0.0112 &
  0.0017 &
  0.0115 &
  0.0044 &
  {\ul 0.0145} &
  0.0051 &
  0.0007 &
  {\ul 0.0069} &
  0.0086 &
  0.0047 &
  0.0102 &
  \textbf{0.0101} &
  \textbf{0.0218} \\
 &
  \multicolumn{1}{c|}{SimGCL} &
  0.0058 &
  0.0186 &
  {\ul 0.0119} &
  0.0214 &
  0.0067 &
  \textbf{0.0367} &
  0.0055 &
  0.0172 &
  0.0109 &
  0.0224 &
  0.0095 &
  0.0248 &
  0.0089 &
  0.0209 &
  0.0108 &
  0.0199 &
  \textbf{0.0159} &
  {\ul 0.0264} \\
 &
  \multicolumn{1}{c|}{XSimGCL} &
  0.0059 &
  0.0127 &
  0.0027 &
  0.0147 &
  0.0030 &
  0.0121 &
  {\ul 0.0072} &
  0.0124 &
  0.0066 &
  0.0105 &
  0.0066 &
  0.0148 &
  0.0055 &
  {\ul 0.0154} &
  0.0067 &
  0.0135 &
  \textbf{0.0100} &
  \textbf{0.0158} \\ \midrule
\multirow{5}{*}{Epinions} &
  \multicolumn{1}{c|}{LightGCN} &
  0.0042 &
  0.0187 &
  0.0023 &
  {\ul 0.0362} &
  0.0011 &
  0.0106 &
  0.0091 &
  0.0312 &
  0.0033 &
  0.0235 &
  0.0074 &
  0.0342 &
  0.0087 &
  0.0303 &
  {\ul 0.0093} &
  0.0256 &
  \textbf{0.0095} &
  \textbf{0.0380} \\
 &
  \multicolumn{1}{c|}{SSL4Rec} &
  0.0082 &
  0.0269 &
  0.0055 &
  0.0198 &
  0.0095 &
  0.0289 &
  0.0325 &
  0.0389 &
  0.0278 &
  0.0306 &
  0.0499 &
  0.0569 &
  {\ul 0.0512} &
  {\ul 0.0582} &
  0.0465 &
  0.0542 &
  \textbf{0.0556} &
  \textbf{0.0658} \\
 &
  \multicolumn{1}{c|}{SGL} &
  0.0151 &
  0.0560 &
  0.0039 &
  0.0220 &
  0.0075 &
  0.0393 &
  0.0186 &
  0.0478 &
  0.0164 &
  0.0356 &
  0.0249 &
  0.0642 &
  {\ul 0.0289} &
  {\ul 0.0753} &
  0.0222 &
  0.0579 &
  \textbf{0.0415} &
  \textbf{0.1223} \\
 &
  \multicolumn{1}{c|}{SimGCL} &
  0.0117 &
  0.0560 &
  0.0037 &
  0.0194 &
  0.0061 &
  0.0328 &
  0.0297 &
  0.0219 &
  0.0288 &
  0.0465 &
  {\ul 0.0311} &
  {\ul 0.0716} &
  0.0310 &
  0.0710 &
  0.0299 &
  0.0345 &
  \textbf{0.0342} &
  \textbf{0.0857} \\
 &
  \multicolumn{1}{c|}{XSimGCL} &
  0.0161 &
  0.0826 &
  0.0048 &
  0.0398 &
  0.0089 &
  0.0499 &
  0.0159 &
  0.0468 &
  0.0244 &
  0.0568 &
  0.0361 &
  0.0941 &
  0.0368 &
  0.0962 &
  {\ul 0.0458} &
  {\ul 0.1262} &
  \textbf{0.0470} &
  \textbf{0.1490} \\ \midrule 
  \multirow{5}{*}{ML-1M} &
  \multicolumn{1}{c|}{LightGCN} &
  0.0020 &
  0.0091 &
  0.0023 &
  0.0101 &
  \textbf{0.0025} &
  {\ul 0.0102} &
  0.0022 &
  0.0098 &
  0.0021 &
  0.0088 &
  0.0021 &
  0.0056 &
  0.0023 &
  0.0068 &
  0.0020 &
  0.0063 &
  {\ul 0.0024} &
  \textbf{0.0111} \\
 &
  \multicolumn{1}{c|}{SSL4Rec} &
  0.0038 &
  0.0100 &
  0.0039 &
  {\ul 0.0111} &
  0.0036 &
  0.0086 &
  0.0029 &
  0.0078 &
  0.0033 &
  0.0108 &
  0.0030 &
  0.0049 &
  0.0041 &
  0.0089 &
  {\ul 0.0043} &
  0.0099 &
  \textbf{0.0046} &
  \textbf{0.0137} \\
 &
  \multicolumn{1}{c|}{SGL} &
  0.0021 &
  0.0102 &
  0.0029 &
  0.0108 &
  0.0033 &
  {\ul 0.0118} &
  0.0030 &
  0.0111 &
  0.0036 &
  0.0106 &
  {\ul 0.0041} &
  0.0103 &
  0.0031 &
  0.0100 &
  0.0035 &
  0.0111 &
  \textbf{0.0052} &
  \textbf{0.0120} \\
 &
  \multicolumn{1}{c|}{SimGCL} &
  0.0040 &
  0.0138 &
  {\ul 0.0072} &
  \textbf{0.0174} &
  0.0039 &
  0.0134 &
  0.0048 &
  0.0098 &
  0.0048 &
  0.0112 &
  0.0036 &
  0.0116 &
  0.0033 &
  0.0112 &
  0.0055 &
  0.0147 &
  \textbf{0.0084} &
  {\ul 0.0142} \\
 &
  \multicolumn{1}{c|}{XSimGCL} &
  0.0034 &
  {\ul 0.0132} &
  0.0033 &
  0.0110 &
  0.0035 &
  0.0130 &
  0.0034 &
  0.0111 &
  0.0032 &
  0.0109 &
  0.0033 &
  0.0112 &
  {\ul 0.0036} &
  0.0117 &
  0.0032 &
  0.0101 &
  \textbf{0.0050} &
  \textbf{0.0138} \\ \midrule
\multirow{5}{*}{Yelp} &
  \multicolumn{1}{c|}{LightGCN} &
  0.0188 &
  0.0114 &
  0.0205 &
  {\ul 0.0122} &
  0.0 &
  0.0 &
  0.0158 &
  0.0094 &
  0.0132 &
  0.0111 &
  {\ul 0.0210} &
  0.0083 &
  0.0125 &
  0.0041 &
  0.0112 &
  0.0038 &
  \textbf{0.0262} &
  \textbf{0.0138} \\
 &
  \multicolumn{1}{c|}{SSL4Rec} &
  0.1563 &
  0.0153 &
  0.1500 &
  0.0540 &
  {\ul 0.2061} &
  0.0190 &
  0.1118 &
  0.0018 &
  0.1886 &
  0.0645 &
  0.0812 &
  0.0276 &
  0.1938 &
  {\ul 0.0716} &
  0.1311 &
  0.0447 &
  \textbf{0.2563} &
  \textbf{0.0932} \\
 &
  \multicolumn{1}{c|}{SGL} &
  0.0937 &
  0.0318 &
  0.1438 &
  0.0494 &
  {\ul 0.1561} &
  {\ul 0.0525} &
  0.0954 &
  0.0348 &
  0.1005 &
  0.0426 &
  0.0268 &
  0.0123 &
  0.1000 &
  0.0337 &
  0.1135 &
  0.0467 &
  \textbf{0.2001} &
  \textbf{0.0674} \\
 &
  \multicolumn{1}{c|}{SimGCL}  &
  0.0250 &
  0.0122 &
  0.0237 &
  0.0141 &
  0.0187 &
  0.0163 &
  0.0258 &
  {\ul 0.0187} &
  0.0232 &
  0.0156 &
  0.0225 &
  {\textbf{0.0209}} &
  {\ul 0.0262} &
  0.0180 &
  0.0198 &
  0.0164 &
  \textbf{0.0375} &
  0.0119 \\
 &
  \multicolumn{1}{c|}{XSimGCL} &
  0.0197 &
  0.0165 &
  0.0270 &
  0.0132 &
  0.0125 &
  0.0143 &
  0.0226 &
  0.0186 &
  0.0202 &
  0.0145 &
  0.0287 &
  {\ul 0.0264} &
  0.0165 &
  0.0043 &
  {\ul0.0305} &
  0.0113 &
  \textbf{0.0365} &
  \textbf{0.0282} \\ \midrule
\end{tabular}}
\label{performance comparison}
\end{table*}

\section{Experiments}

\subsection{Experimental Settings}
\stitle{Datasets.} Four public datasets with different scales: DouBan \cite{32Zhao}, Epinions \cite{58pan2020learning}, ML-1M \cite{33FangYGL18}, and Yelp \cite{09yu2022graph}, are used in our experiments. The dataset statistics are shown in Appendix \ref{Statistics of Dataset}.

\stitle{Evaluation Protocol.} We split the datasets into three parts (training set, validation set, and test set) in a ratio of 7:1:2. The poisoning data only affects the model training process; therefore, we exclusively utilize the training set data as the known data. Two metrics are used for measuring attack efficiency, i.e., Hit Ratio@50 and NDCG@50, which calculate the frequency and ranking of the target items in recommendation lists. 
% The formula of Hit Ratio@K and NDCG@K can be seen in Appendix \ref{Evaluation Metrics}. 
Each experiment in this section is conducted 10 times, and then we report the average results.

\stitle{Recommendation Methods.} We choose LightGCN \cite{07he2020lightgcn} as the base model and five recent CL-based recommendation models as the subjects of study:

\begin{itemize}[leftmargin=12pt]
\item SSL4Rec \cite{55yao2021self} introduces a contrastive objective to enhance item representation by utilizing the inherent relationships among item features.
\item SGL \cite{08wu2021self} adopts node/edge dropout to augment graph data and conducts CL between representations learned over different augmentations to enhance recommendation.
\item SimGCL \cite{09yu2022graph} simplifies data augmentations in CL and optimizes representations towards more uniformity.
\item XSimGCL \cite{25yu2022xsimgcl} further streamlines the forward/backward pass in SimGCL for more effective and efficient CL.
\end{itemize}

\stitle{Attack Comparison Methods.} To examine the effectiveness of the proposed attack method, we compare it with some representative poisoning attacks.
\begin{itemize}[leftmargin=12pt]
\item Random Attack \cite{10lam2004shilling} randomly interacts with available items to construct attacks with minimum expenses. 
\item Bandwagon Attack \cite{21li2016data} chooses to interact with popular items to expose target items to more users.
\item AUSH \cite{22lin2020attacking} leverages GANs \cite{34creswell2018generative} to generate malicious users based on real user profiles.
\item DLAttack \cite{59huang2021data} formulates the attack as an optimization problem in the deep learning-based recommendation scenario.
\item PoisonRec \cite{66song2020poisonrec} leverages reinforcement learning to dynamically generate interactive behavior for fake user profiles.
\item FedRecAttack \cite{20rong2022fedrecattack} utilizes a bi-level optimization approach to generate the interactions between malicious users and items. 
\item A-hum \cite{23rong2022poisoning} identifies hard users who consider the target items as negative samples, and manipulates malicious interactions to elevate these users' preferences for the targeted items.
\item GSPAttack \cite{67nguyen2023poisoning} targets graph neural network-based recommendation by exploiting the vulnerabilities of graph propagation, implementing attacks through the generation of fake users and user-item interaction data.
\end{itemize}

It should be noted that FedRecAttack and A-hum are designed under model poisoning setting which manipulates continuous gradients in federated RS. To ensure fairness, we have adapted FedRecAttack and A-hum to data poisoning that manipulates discrete interactions in Centralized RS.

% \stitle{Parameter Settings.} Please see Appendix \ref{Parameter Settings}.

\subsection{Attack Performance Comparison}
We begin by comparing CLeaR with existing attack methods on four distinct datasets. The results are presented in Table \ref{performance comparison}. In the table, values shown in bold represent the best performing indices, while the runner-ups are indicated with underlines. Upon examining Table \ref{performance comparison}, we can derive the following observations and conclusions:

\begin{itemize} [leftmargin=*]
    \item The proposed CLeaR effectively enhances the likelihood of target items being included in the recommendation list generated by CL-based recommendation models. CLeaR demonstrates the best performance among all existing methods in most cases. These performances can be attributed to CLeaR's consideration of representation spectra. Throughout the attack process, CLeaR promotes a smooth spectra distribution, while other attack methods fail to take this factor into account in their approach.
    \item Conventional attack methods, such as Random-Attack and Bandwagon-Attack, typically fall short of the efficacy demonstrated by learning-based attack approaches in most instances.
    \item CLeaR proves to be effective not just within CL-based RS but also in those non-CL counterparts. This strategic attack focus extends CLeaR's applicability and increases its potential threat across various recommendation methods.
\end{itemize}

% \vspace{-10pt}
\subsection{Impact of Attack Size}
We assess the performance of CLeaR in situations with a growing number of malicious users. To maintain attack practicality, we limit the attack size to specific percentages, specifically [1\%, 2\%, 3\%, 4\%, 5\%]. Figure \ref{attacksize} displays the experimental results on two datasets, and other datasets exhibit patterns analogous to those identified in these two datasets. Upon analyzing the data presented in Table \ref{performance comparison}, the following insights and conclusions can be extrapolated:

\begin{figure}[h]
  \centering
  \includegraphics[width=1\linewidth]{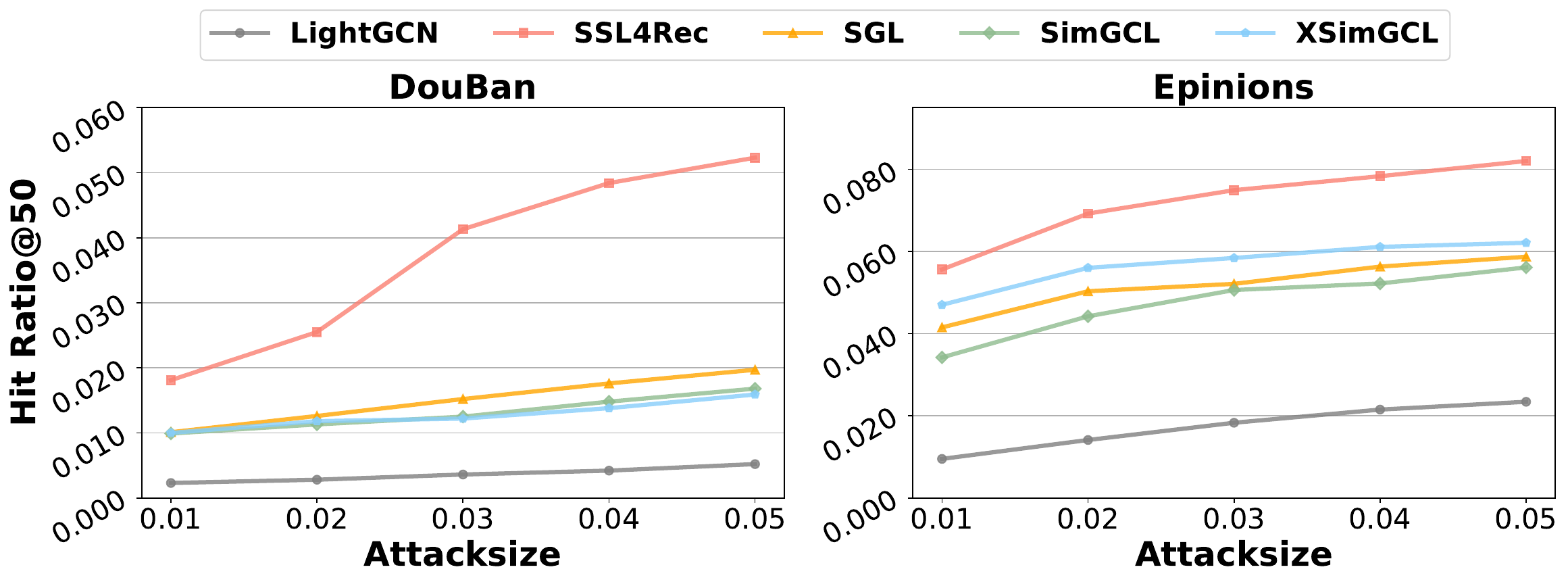}
  \vspace{-10pt}
  \caption{Attack performance w.r.t. attack size.}
  \vspace{-10pt}
\label{attacksize}
\end{figure}

\begin{itemize} [leftmargin=*]
\item
CL-based recommendation models indicate variable degrees of sensitivity to different attack sizes. Specifically, SSL4Rec exhibits a pronounced increase in vulnerability, while models like SGL, SimGCL, and XSimGCL show a moderate, upward trend in their hit ratios corresponding to incremental increases in attack size. Parallel observations are evident in the Epinions dataset.
\item
LightGCN demonstrates a relatively stable performance over the range of attack sizes, exhibiting robustness that renders it less susceptible to attacks compared to other CL-based recommendation methods. This observation aligns with the initial findings presented earlier in the paper.
\item
CLeaR consistently exhibits upward trends in performance as the attack size increases across all cases. It's crucial to highlight that these upward trends have not reached a plateau, indicating that CLeaR's full attack potential remains untapped. This highlights the need to closely monitor CLeaR for potential threats. 
\end{itemize}

\subsection{The Effect of Two Promotion Objectives}
In our experiments, we assess the effect of the dispersion promotion objective and rank promotion objective on CLeaR. Table \ref{effect} illustrates the efficacy of two objectives as individual and combined attack strategies. $\text{CLeaR}_{D}$, $\text{CLeaR}_{R}$, and $\text{CLeaR}_{D+R}$ denote the CLeaR attack method configured with solely the dispersion promotion objective, exclusively the rank promotion objective, and a combination of both objectives, respectively. From Table \ref{effect}, we can find the observations as follows:
\begin{itemize} [leftmargin=*]
\item 
Without any attack (NoneAttack), the Hit Ratio@50 is notably low across all the models, indicating a minimal presence of target items in the absence of manipulation. 
\item
The implementations of the dispersion promotion ($\text{CLeaR}_{D}$) and rank promotion ($\text{CLeaR}_{R}$) as a standalone attack strategy result in a discernible improvement in Hit Ratio@50 compared to the NoneAttack scenario. This suggests that $\text{CLeaR}_{D}$ passively reduces the difficulty for an attacker to place target items within the top recommendations successfully, and $\text{CLeaR}_{R}$ actively promotes target items closer to the user in the recommendation list. The combined use of dispersion promotion objective and rank promotion objective ($\text{CLeaR}_{D+R}$) demonstrates the most substantial increase in Hit Ratio@50, highlighting a synergistic effect where the conjunction of balance and rank promotion objectives culminates in an optimized attack impact.
\end{itemize}

\begin{table}[t]
\caption{The effect of two promotion objectives of CLeaR on DouBan and Epinions (measured by Hit Ratio@50).}

\centering
\resizebox{0.43\textwidth}{!}{
\begin{tabular}{@{}c|c|c|c|c|c@{}}
\toprule
\textbf{Dataset} & \textbf{Cases} & \textbf{SSL4Rec} & \textbf{SGL} & \textbf{SimGCL} & \textbf{XSimGCL} \\ \midrule
\multirow{4}{*}{DouBan}   & NoneAttack & 0.0             & 0.0001          & 0.0002          & 0.0001          \\
                          & $\text{CLeaR}_{D}$     & 0.0057          & 0.0047          & 0.0051          & 0.0035          \\
                          & $\text{CLeaR}_{R}$    & 0.0136          & 0.0051          & 0.0076          & 0.0064          \\
                          & $\text{CLeaR}_{D+R}$      & \textbf{0.0181} & \textbf{0.0101} & \textbf{0.0099} & \textbf{0.0100} \\ \midrule
\multirow{4}{*}{Epinions} & NoneAttack & 0.0             & 0.0001          & 0.0             & 0.0             \\
                          & $\text{CLeaR}_{D}$     & 0.0158          & 0.0168          & 0.0094          & 0.0245          \\
                          & $\text{CLeaR}_{R}$    & 0.0359          & 0.0219          & 0.0303          & 0.0301          \\
                          & $\text{CLeaR}_{D+R}$      & \textbf{0.0556} & \textbf{0.0415} & \textbf{0.0342} & \textbf{0.0470} \\ \bottomrule
\end{tabular}}
\label{effect}
\end{table}

\subsection{\textbf{Attack Analysis on More Basic Encoders}}
\label{basic recommendation encoders}
From the data results and analysis in Table \ref{performance comparison}, we can conclude that CLeaR not only works on CL-based recommendation methods. Therefore, we design additional experiments to explore whether CLeaR can have a stronger attack effect on a wider range of basic recommendation encoders. Specifically, we select GMF \cite{60koren2009matrix} and NGCF \cite{61wang2019neural} as the basic recommendation encoders to be attacked. The results are shown in Table \ref{transferity}, where CLeaR still shows the strongest attack. This advantage is most obvious when attacking NGCF on the Epinions data set. These experimental results show that CLeaR's consideration of promoting smooth spectral distribution can indeed be widely used to attack various recommendation systems. This has sounded the alarm for us, and we should pay attention to this feature when designing recommendation algorithms.

\begin{table}[h]
\caption{Comparison of different attack methods with basic encoders (measured by Hit Ratio@50).} 

\centering
\resizebox{0.36\textwidth}{!}{
\begin{tabular}{@{}c|cc|cc@{}}
\toprule
\textbf{Dataset} & \multicolumn{2}{c|}{\textbf{DouBan}}                   & \multicolumn{2}{c}{\textbf{Epinions}} \\ \midrule
Models       & \multicolumn{1}{c|}{GMF}    & NGCF   & \multicolumn{1}{c|}{GMF}    & NGCF   \\ \midrule
RandomAttack & \multicolumn{1}{c|}{0.0017} & 0.0019 & \multicolumn{1}{c|}{0.0032} & 0.0037 \\
AUSH         & \multicolumn{1}{c|}{0.0019} & 0.0021 & \multicolumn{1}{c|}{0.0040} & 0.0063 \\
DLAttack     & \multicolumn{1}{c|}{0.0021} & 0.0021 & \multicolumn{1}{c|}{0.0039} & 0.0058 \\
FedRecAttack & \multicolumn{1}{c|}{0.0017} & 0.0021 & \multicolumn{1}{c|}{0.0059} & 0.0068 \\
A\_hum        & \multicolumn{1}{c|}{0.0016} & 0.0020 & \multicolumn{1}{c|}{0.0060} & 0.0073 \\ \midrule
CLeaR            & \multicolumn{1}{c|}{\textbf{0.0023}} & \textbf{0.0023} & \multicolumn{1}{c|}{\textbf{0.0066}}   & \textbf{0.0100}   \\ \bottomrule
\end{tabular}}
\label{transferity}
\end{table}

\subsection{\textbf{Attack Analysis on Black-box Setting}}
\label{black-box setting}
In this experiment, we demonstrated the adaptability of our CLeaR to black-box attacks by using surrogate models on the DouBan dataset. Specificially, all victim models utilized LightGCN as the base encoder, and we selected NGCF and LightGCN as surrogate models, respectively. To simulate a realistic attack scenario, we train both the victim models and surrogate models independently, even when they share the same model structure. As can be seen from table \ref{tab:comparison}, our attack method exceeded the performance of existing baseline methods in most cases.This demonstrates that our approach adapts effectively to black-box settings when using a general recommendation model as the surrogate model. Furthermore, our experiments also led to two noteworthy findings:
\begin{itemize} [leftmargin=*]
\item
Our method shows superior attack performance when the model structure of the surrogate model is similar to that of the victim model, compared to when it differs. This suggests that if we can obtain prior knowledge about the model structure of the victim model, it can be effectively incorporated into our surrogate model training process to facilitate a more effective attack;
\item 
While possessing knowledge of the victim model’s structure aids the attack process, it does not achieve the effectiveness level of a white-box attack. This is due to the fact that, although the models have an identical structure, they are trained independently, leading to variations in the trained model parameters.
\end{itemize}

\begin{table}[h]
\caption{Comparison between the black-box baseline and CLeaR variants (measured by Hit Ratio@50).}

\centering
\resizebox{0.45\textwidth}{!}{
\begin{tabular}{l|c|c|c|c}
\hline
\textbf{Models}    & \textbf{PoisonRec} & \textbf{CLeaR} & \textbf{CLeaR} & \textbf{CLeaR} \\
                   &                    & \textbf{-LightGCN} & \textbf{-NGCF} & \textbf{-White} \\
\hline
LightGCN  & 0.0016    & 0.0021         & 0.0019     & \textbf{0.0023}      \\
SSL4Rec   & 0.0146    & 0.0171         & 0.0166     & \textbf{0.0181}      \\
SGL       & 0.0044    & 0.0089         & 0.0084     & \textbf{0.0101}      \\
SimGCL    & 0.0109    & 0.0132         & 0.0128     & \textbf{0.0159}      \\
XSimGCL   & 0.0066    & 0.0082         & 0.0077     & \textbf{0.0100}      \\
\hline
\end{tabular}}
\label{tab:comparison}
\end{table}

\vspace{-10pt}
\subsection{\textbf{Sensitivity of Hyperparameters}}
\label{Sensitivity of Hyperparameters}
We examine the sensitivity of the CLeaR model to the hyperparameter $\alpha$ on DouBan and Epinions datasets. We vary $\alpha$ within a set [0, 0.01, 0.1, 0.5, 1, 5, 10]. The results are illustrated in Fig. \ref{parameters}, and we can derive the following observations: CLeaR indicates consistent stability in performance across all datasets as $\alpha$ varies. LightGCN on the DouBan dataset shows a consistent increase in the Hit Ratio@50 as $\alpha$ increases, reaching its peak at $\alpha$ =0.5. SSL4Rec displays an initial increase and then plateaus, maintaining performance across higher $\alpha$ values. Both SGL and SimGCL experience a modest rise before reaching stability, while XSimGCL displays a gentle upward trend but largely maintains a steady performance. A similar pattern is observed from the results conducted on Epinions.

% \begin{itemize} [leftmargin=*]
% \item
% CLeaR indicates consistent stability in performance across all datasets as $\alpha$ varies. LightGCN on the DouBan dataset shows a consistent increase in the Hit Ratio@50 as $\alpha$ increases, reaching its peak at $\alpha$ =0.5. SSL4Rec displays an initial increase and then plateaus, maintaining performance across higher $\alpha$ values. Both SGL and SimGCL experience a modest rise before reaching stability, while XSimGCL displays a gentle upward trend but largely maintains a steady performance. A similar pattern is observed from the results conducted on Epinions.
% \item
% The experimental results affirm the efficacy of the dispersion promotion objective within CLeaR. This is particularly evident when the $\alpha$=0, where CLeaR demonstrates the lowest effective attack impact, underscoring the significance of the dispersion promotion component in the overall attack strategy.
% \end{itemize}

\subsection{Analysis of Time Complexity }
In this part, we report the real running time of compared methods for a one-time attack on DouBan dataset. The results in Table ~\ref{runtime} are collected on an Intel(R) Core(TM) i9-10900X CPU and a GeForce RTX 3090Ti GPU. Table ~\ref{runtime} presents the running time in seconds for a one-time attack on the DouBan dataset across various attack methods and recommender system models: LightGCN, SGL, SimGCL, and XSimGCL. The observation can be found as follows: 
\begin{itemize} [leftmargin=*]
\item
Among all evaluated models, RandomAttack consistently exhibits the lowest execution times, primarily due to its training-free and heuristic-based methodology. CLeaR, on the other hand, presents moderate execution times when compared to other attacks, but is significantly more time-consuming than RandomAttack. This indicates that CLeaR strikes a computational balance, effectively navigating the trade-off between operational efficiency and algorithmic complexity.
\item
RandomAttack and AUSH maintain stable execution times across models, reflecting their model-independent nature. In contrast, other strategies exhibit variable time costs contingent on the recommendation model in use. This variability is attributed to the necessity for these methods to train attack models tailored to specific recommendation model structures, in contrast to the model-agnostic approach employed by RandomAttack and AUSH.
\end{itemize}

\begin{table}[h]
\caption{Running time (s) for a one-time attack on DouBan.}

\centering
\resizebox{0.48\textwidth}{!}{
\begin{tabular}{c|c|c|c|c|c}
\midrule
\textbf{Models} & \textbf{LightGCN} & \textbf{SSL4Rec} & \textbf{SGL}    & \textbf{SimGCL} & \textbf{XSimGCL} \\ \midrule
RandomAttack & 7.7167 & 7.5945 & 7.3935 & 7.4895 & 7.6484 \\
AUSH         & 1340.6 & 1348.1 & 1379.3 & 1356.3 & 1378.1 \\
DLAttack     & 1736.3 & 4537.2 & 7654.1 & 5671.0 & 4876.2 \\
FedRecAttack & 777.36 & 899.82 & 1222.6 & 957.94 & 929.64 \\
A\_hum        & 768.28 & 884.62 & 1273.4 & 950.75 & 912.15 \\ \midrule
CLeaR           & \textbf{345.43}   & \textbf{428.64}  & \textbf{885.43} & \textbf{735.64} & \textbf{454.54}  \\ \midrule
\end{tabular}}
\label{runtime}
\end{table}

% \stitle{Extended Experimental Results and Analyses.} Additional experiments detailing the sensitivity of hyperparameters and the impact of CLeaR on more basic recommendation encoders can be found in the Appendix \ref{Sensitivity of Hyperparameters}.

\section{RELATED WORK}
\stitle{CL-Based Recommendation.} 
CL-based recommendation methods are specifically designed to maximize the agreement between positive pair representations while minimizing the similarity between negative pair representations. Through this process, the recommendation model can learn the essential information from the user-item interactions. 

For instance, $S^{3}$-Rec \cite{38zhou2020s3},  CL4SRec \cite{39xie2022contrastive}, SSL4Rec \cite{55yao2021self} and DuoRec \cite{40qiu2022contrastive} introduce CL into sequence augmentations by randomly masking attributes and items, emphasizing the consistency between different augmentations. During the same period, CL is also introduced to various graph-based recommendation scenarios. HHGR \cite{41zhang2021double} proposes a double-scale augmentation approach for group recommendation. CrossCBR \cite{42ma2022crosscbr} has explored the application of CL in cross-domain recommendation. NCL \cite{43lin2022improving} introduces a prototypical contrastive objective that captures the correlations between a user/item and its context. SEPT \cite{44yu2021socially} and COTREC \cite{xia2021self} go a step further and propose semi-supervised learning approaches for social/session-based recommendation. SGL \cite{08wu2021self} generates two additional views of graphs through node/edge dropout for CL. In a similar way, SimGCL \cite{09yu2022graph} directly creates an additional view at the representation level. To streamline the propagation process, XSimGCL \cite{25yu2022xsimgcl} simplifies the forward/backward pass of SimGCL for both recommendation and contrastive tasks.

\stitle{Poisoning Attacks.}
In the past, traditional attacks on recommender systems relied on heuristic manipulations of user behavior data. Random Attack \cite{10lam2004shilling} involved users randomly rating items, while Bandwagon Attack \cite{21li2016data} focused on interacting with popular items to boost the target item's visibility. However, as recommender systems have become more sophisticated and robust, these conventional attack methods have become less effective. In response, researchers have turned to machine learning techniques to develop more advanced attacks \cite{62wang2021fast, 64yuan2023manipulating}.

AUSH \cite{22lin2020attacking}, GOAT \cite{45wu2021ready}, and TrialAttack \cite{47wu2021triple} use generative adversarial networks \cite{34creswell2018generative} to create fake user profiles based on existing real user configurations. This technique employs a discriminator to differentiate between real and fake users, providing guidance to the attacker. PoisonRec \cite{66song2020poisonrec}, LOKI \cite{46zhang2020practical} and KGAttack \cite{48chen2022knowledge} propose reinforcement learning-based attack strategies to generate interactive behavior sequences for fake user profiles. All the aforementioned methods are typical black-box attacks, assuming that the attacker has no knowledge of the internal structure or parameter information of the recommendation model. However, in certain attack scenarios, the attacker may possess complete or partial information about the recommendation model. For instance, DLAttack \cite{59huang2021data} assumes that attacker know all the details of recommendation models and explore poisoning attack in the deep learning-based recommendation scenario. FedRecAttack \cite{20rong2022fedrecattack} utilizes a bi-level optimization approach to optimize interactions between malicious users and items. Similarly, A-hum \cite{23rong2022poisoning} identifies "hard users" who consider the target items as negative samples. By manipulating these hard users through specific processes, the attacker can ensure the inclusion of the target items in their recommendation lists. 

% It is noteworthy that existing attacks have not specifically targeted CL-based recommendation systems. Given the extensive research on CL-based recommendation, we have investigated their vulnerabilities from the perspective of CL loss and demonstrated improved attack performance through the manipulation of representation's spectral values.

\section{CONCLUSION}
In this study, we identify the susceptibility of CL-based recommendation methods to poisoning attacks that specifically promote target items. We also investigate the underlying reasons for this sensitivity and demonstrate that the global dispersion of embedding distributions induced by the CL loss is the core factor. Additionally, we introduce a novel attack approach called CLeaR, which manipulates the spectral value distribution of representations to be smoother and simultaneously optimizes the visibility of target items in users' preference lists. The extensive experiments conducted on four real-world datasets demonstrate the potential threats. Our future research endeavors will concentrate on developing more nuanced defense approaches within CL-based recommender systems to tackle the specific challenges associated with poisoning attacks. 
%A viable defense approach involves clustering items into distinct groups and assigning different contrastive loss weights accordingly. This re-weighting strategy allows for the fine-tuning of CL loss's impact, enhancing recommender systems' robustness.

\begin{acks}
This research is funded by the National Natural Science Foundation of China (Grant No. 62176028), the Australian Research Council through the Future Fellowship (Grant No. FT210100624), Discovery Project (Grants No. DP240101108), Industrial Transformation Training Centre (Grant No. IC200100022), and the Science and Technology Research Program of Chongqing Municipal Education Commission (Grant No. KJZD-K202204402). Additionally, this work is supported by the ARC Training Centre for Information Resilience (CIRES).
\end{acks}

\newpage
\bibliographystyle{ACM-Reference-Format}
\bibliography{CLeaR}

\appendix
% \appendixpage
% \clearpage

\section{APPENDIX}
\subsection{Visualization Experiment of Representation Distributions
Under Attacks}
\label{Visualization Experiment}

We conducted experiments for visualizing the representation distributions learned by recommendation models with and without CL when subjected to a poisoning attack. The experimental setup employed is as follows: We use Random-Attack as the chosen attack approach, and for each malicious user, the number of interactions is set to match the average number of user interactions. To ensure clear and concise experimental results, we employed a systematic sampling strategy for each dataset. Initially, items were ranked based on their popularity, and we sampled 500 popular items from the top 20\% group, along with 500 cold items and 5 target items from the remaining items. This sampling approach was chosen to accommodate the presence of a long-tail distribution (also named power-law distribution) in item popularity. Additionally, we randomly selected 500 users for the experiment. Then, we used t-SNE \cite{17van2008visualizing} to map the optimal representations to a two-dimensional space. 

\begin{figure*}[]
  \centering
    \label{hitcomparisonml100k}
    \includegraphics[width=1\textwidth]{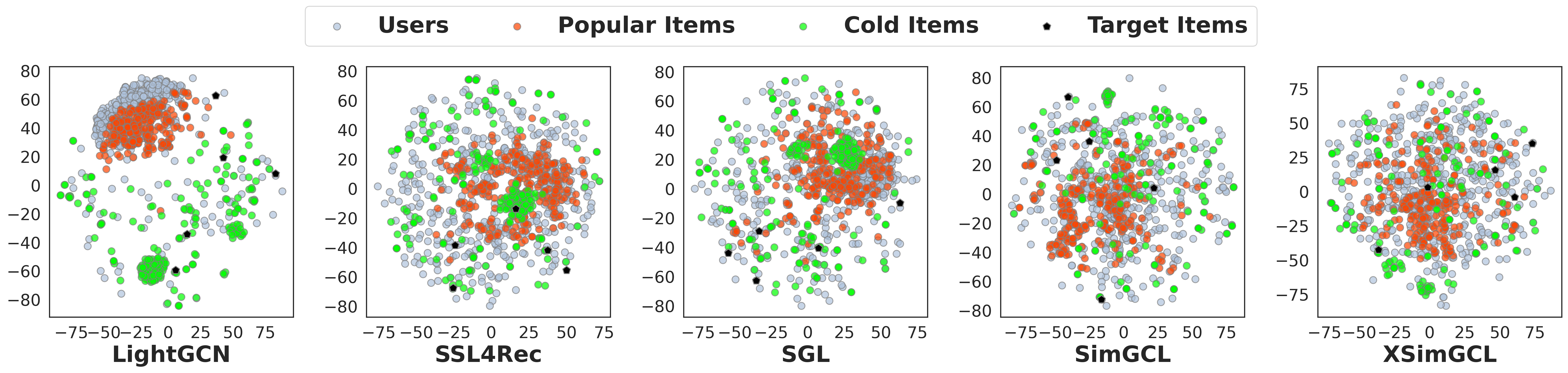}
  \caption{Representation distribution on DouBan dataset under poisoning attacks (Random-Attack).} \vspace{-1em}
  \label{embeding}
\end{figure*}

Our experimental results were executed on a selection of prevalent datasets, including Epinions \cite{58pan2020learning} and DouBan \cite{32Zhao}. Figure~\ref{Epinions_embeding} and Figure~\ref{embeding} demonstrate two patterns: local clustering in the absence of CL and global dispersion with CL. Local clustering aggregates popular items and users, isolating cold items. In contrast, global dispersion achieves a balanced spread across the vector space, reducing the distance between popular and cold items.

% \begin{figure*}[]
%   \centering
%   \subfigure[Representation distribution on DouBan under poisoning attacks.]{
%     \label{hitcomparisonml100k}
%     \includegraphics[width=1\textwidth]{douban_embeding.png}
%   }
%   \subfigure[Representation distribution on ML-1M under poisoning attacks.]{
%     \label{hitcomparisonlastfm}
%     \includegraphics[width=1\textwidth]{ml1m_embeding.png}
%   }
%   \caption{Representation distribution on DouBan and ML-1M datasets under poisoning attacks (Random-Attack).} \vspace{-1em}
%   \label{embeding}
% \end{figure*}

\vspace{-0.5em}
\subsection{Proof of Proposition 1}
\label{Proof of Proposition 1}
\stitle{Proposition 1}. \textit{Give the representations $\textbf{Z}^\prime$ \textit{and} $\textbf{Z}^{\prime\prime}$ which are learned on augmented views} \textit{and the corresponding singular values } $\mathbf{\Sigma}' = diag(\mathbf{\sigma}'_{1}, ..., \mathbf{\sigma}'_{d})$ \textit{and} $\mathbf{\Sigma}'' = diag(\mathbf{\sigma}''_{1}, ..., \mathbf{\sigma}''_{d})$, \textit{an upper bound of the CL loss is given by:}
\begin{equation}
\begin{aligned}
\mathcal{L}_{cl} < N \max_{j} \sigma'_{j}  \sigma''_{j} -\sum_{i} \sigma'_{i} \sigma''_{i} + N \log N.
\end{aligned}
\end{equation}

\stitle{Proof.} For simplification, we first set $\tau=1$ in Equation \eqref{cllosstransformation}, and we can get that:
\begin{equation}
\begin{aligned}
\mathcal{L}_{cl} &= - \sum_{p \in \mathcal{(U\cup I)}} \overline{\mathbf{z}}'^T_p \overline{\mathbf{z}}''_p + \sum_{p \in \mathcal{(U\cup I)}} \log  \sum_{n \in \mathcal{(U\cup I)}} \exp (\overline{\mathbf{z}}'^T_p \overline{\mathbf{z}}''_{n})\\
&< - \sum_{p \in \mathcal{(U\cup I)}} \overline{\mathbf{z}}'^T_p \overline{\mathbf{z}}''_p +  \sum_{p \in \mathcal{(U\cup I)}} \log (N \max_{n \in \mathcal{(U\cup I)}} \exp(\overline{\mathbf{z}}'^T_p \overline{\mathbf{z}}''_{n}))\\
&< - \sum_{p \in \mathcal{(U\cup I)}} \overline{\mathbf{z}}'^T_p \overline{\mathbf{z}}''_p + \sum_{p \in \mathcal{(U\cup I)}}( \log N + \max_{n \in \mathcal{(U\cup I)}} \overline{\mathbf{z}}'^T_n \overline{\mathbf{z}}''_{n}).
\end{aligned}
\end{equation}

In the provided equation, the transformation from the first to the second row applies the log-sum-exp trick, consolidating all $\overline{\mathbf{z}}'^T_p \overline{\mathbf{z}}''_{n}$ terms to their maximum within the set, thus simplifying the computation. Proceeding to the third row, drawing inspiration from the findings of \cite{19wang2020understanding, 49wang2022towards}, we suppose a near alignment of the two augmented representations. This implies that the maximum similarity between any negative pair $\overline{\mathbf{z}}'^T_p \overline{\mathbf{z}}''_{n}$ is less than the maximum similarity within the augmented representations themselves $\overline{\mathbf{z}}'^T_n \overline{\mathbf{z}}''_n$. Consequently, the optimization of the CL loss can be reformulated as follows: 
\begin{equation}
\begin{aligned}
\mathcal{L}_{cl} &< - \sum_{p \in \mathcal{(U\cup I)}} \overline{\mathbf{z}}'^T_p \overline{\mathbf{z}}''_p + N \max_{n \in \mathcal{(U\cup I)}} (\overline{\mathbf{z}}'^T_n \overline{\mathbf{z}}''_{n}) + N \log N \\
&= - Tr({\textbf{Z}'}^T \textbf{Z}'') + N \max_{n \in \mathcal{(U\cup I)}} (\overline{\mathbf{z}}'^T_n \overline{\mathbf{z}}''_{n}) + N \log N,
\label{tr}
\end{aligned}
\end{equation}
where $Tr(\cdot)$ means the trace of matrix.

Considering ${\textbf{Z}'}$ and ${\textbf{Z}''}$ as augmented representations derived from the original representations, we hypothesize a minimal variance in their respective singular vectors. Therefore, we constrain them to be identical. Following this constraint, we get that $\textbf{Z}'= \mathbf{L} \mathbf{\Sigma}' \mathbf{R}^T$ and $\textbf{Z}''=\mathbf{L} \mathbf{\Sigma}'' \mathbf{R}^T$, leading to:

\vspace{-1em}
\begin{equation}
\begin{aligned}
{\textbf{Z}'}^T \textbf{Z}'' &= \textbf{R} \Sigma'^{T} \textbf{L}^T \textbf{L} \Sigma'' \textbf{R}^T \\
&=\sigma'_{1}\sigma''_{1} \textbf{r}_{1}{\textbf{r}_{1}}^T + \sigma'_{2}\sigma''_{2} \textbf{r}_{2}{\textbf{r}_{2}}^T + \cdots + \sigma'_d\sigma''_{d} \textbf{r}_{d}{\textbf{r}_{d}}^T.
\end{aligned}
\end{equation}

Then, we plug this equation into Equation \eqref{tr}, and due to $\textbf{r}^T_{j} \textbf{r}_{j}=1$, we can get that:
\begin{equation}
\begin{aligned}
\mathcal{L}_{cl} & < -Tr({\textbf{Z}'}^T \textbf{Z}'') + N \max_{n \in \mathcal{(U\cup I)}} (\overline{\mathbf{z}}'^T_n \overline{\mathbf{z}}''_{n}) + N \log N\\
& = -\sum_{i} \sigma'_{i} \sigma''_{i}  + N \max_{j} \sigma'_{j}  \sigma''_{j} \textbf{r}^T_{j} {\textbf{r}_{j}} + N \log N \\
&= N \max_{j} \sigma'_{j}  \sigma''_{j} -\sum_{i} \sigma'_{i} \sigma''_{i} + N \log N.
\end{aligned}
\end{equation}

\subsection{Algorithm}
\label{Algorithm}
We present the attack algorithm of CLeaR in Algorithm \ref{algorithm1}.
\begin{algorithm}[h]
		\caption{The White-Box Algorithm of CLeaR.}
		\label{algorithm1}
         \begin{algorithmic}
         \STATE \textbf{Input:} The recommendation model $f$, malicious users set \\ $\mathcal{U}_{M}$, target items set $\mathcal{I}^{T}$, interactions data $\mathcal{D}$.
         \STATE \textbf{Output:} Malicious Data $\mathcal{D}_{M}$.
         \end{algorithmic}

        \While{not converged}
        {
            \tcp{inner optimization: \\ recommendation phase }

            sample tuples $(u,i,j)$ from the synthetic dataset $\mathcal{D}$ and $\mathcal{D}_{M}$;
            
            calculate ${L}_{rec}$ to obtain optimal user representation $\mathbf{Z}^{*}_{\mathcal{U}}$ and item representation $\mathbf{Z}^{*}_{\mathcal{I}}$;

            \tcp{outer optimization: \\ attack phase}

            \textbf{Dispersion Promotion Objective}: traverse all representations $\mathbf{Z}^{*}_{\mathcal{U}}\cup\mathbf{Z}^{*}_{\mathcal{I}}$, calculate and accumulate every $\frac{||\textbf{Z}\textbf{V}'\textbf{V}'^{T}||}{||\textbf{V}'||_{2}^{2}}$, i.e., calculate $\mathcal{L}_{D}$;

            \textbf{Rank Promotion Objective}: traverse all users $u \in \mathcal{U}$, and find the last item in the recommendation list of user $u$ to calculate $\mathcal{L}_{R}$ ;
            
            update $\mathcal{D}_{M}$, i.e., $\mathcal{D}_{M}$=$\arg\max(\mathcal{L}_{D}$ + $\alpha\mathcal{L}_{R})$;
        }
\end{algorithm} 
\stitle{More Details of Approximate Solution.} The CLeaR utilizes a bi-level optimization framework. In outer optimization, we face the challenge of discrete domains and constraints, as interactions can only take values of 0 or 1. To address this, we adopt a greedy approximation scheme \cite{50zugner2018adversarial}. We treat all interactions as continuous values for optimization and then select the interactions that yield the highest values among them. The number of choices is limited by the maximum number of interactions of malicious users minus the number of target items. The remaining interactions of each malicious user are fixed to interact with the target items. 

\subsection{Experimental Supplements}
\subsubsection{\textbf{Statistics of Dataset}}
\label{Statistics of Dataset}

\begin{table}[h]
\caption{\textbf{Statistics of datasets.}}
\vspace{-10pt}
\small
\centering
\label{datasets}
\begin{tabular}{@{}ccccc@{}}
\toprule
\textbf{Dataset}                                            & \textbf{\#Users} & \textbf{\#Items} & \textbf{\#Interactions} & \textbf{\#Density} \\ \midrule

\begin{tabular}[c]{@{}c@{}}DouBan\end{tabular} & 2,831
          & 36,821
           & 805,611
               & 0.772\%          \\
Epinions                                                    & 75,887          & 75,881          & 508,837               & 0.009\%          \\
ML-1M                                                      & 6,038
          & 3,492
           & 575,281
                & 2.728\%           \\
Yelp                                                    & 31,668          & 38,048           & 1,561,406               & 0.129\%  \\
\bottomrule
\end{tabular} 
\end{table}
% \subsubsection{\textbf{Evaluation Metrics}}
% \label{Evaluation Metrics}

% The formula of Hit Ratio@K is as follows:
% \begin{equation}
%     \begin{aligned}
%         \text{Hit Ratio@K} = \frac{1}{|\mathcal{U}||\mathcal{I}^{T}|} \sum_{u \in \mathcal{U}} {|\mathcal{I}^{T} \cap \mathcal{I}^{rec}_u|},
%     \end{aligned}
% \end{equation}
% where $K$ is the number of items in the recommended list.

% The formula of NDCG@K is as follows:
% \begin{equation}
%     \begin{split}
%     \rm{NDCG@K} &= \frac{1}{|\mathcal{U}|} \sum_{u \in \mathcal{U}} \frac{{\rm DCG}_{u}@K}{{\rm IDCG}@K},
%     \end{split}
%     \label{NDCG}
%     \end{equation}
%     \vspace{-0.5em}
%     \begin{equation}
%     \begin{split}
%     \rm{DCG}_{u}@K &= \sum_{m = 1}^{|\mathcal{I}^{rec}_{u}|} \frac{{rel_m}}{\log_2(m+1)}, 
%     \end{split}
%     \end{equation}
    
%     \begin{equation}
%     \begin{split}
%     \rm{IDCG}@K &= \sum_{m=1}^{|\mathcal{I}^{T}¥|} \frac{1}{\log_2(m+1)},
%     \end{split}
%     \end{equation}
% where $rel_{m}$ means the corresponding item of index $m$ in user $u$'s recommendation list belongs to target items. 

\subsubsection{\textbf{Parameter Settings}} 
\label{Parameter Settings}
As for the specific hyperparameters of the baseline models, we initialize them with the recommended settings in the original papers and then perform grid search around them for the optimal performance. For all attack models, the injected malicious user population is 1\% of the number of normal users, and each malicious user was assigned an interaction count equivalent to the average number of interactions per normal user. As for the general settings, we set $\omega$, batch size, learning rate, embedding size, and the number of LightGCN layers to 0.1, 1024, 0.001, 64, and 2, respectively. Following previous work \cite{10lam2004shilling,21li2016data}, we randomly choose 10 cold items from the bottom 80\% group as target items.

\begin{figure}[h]
  \centering
  \includegraphics[width=1\linewidth]{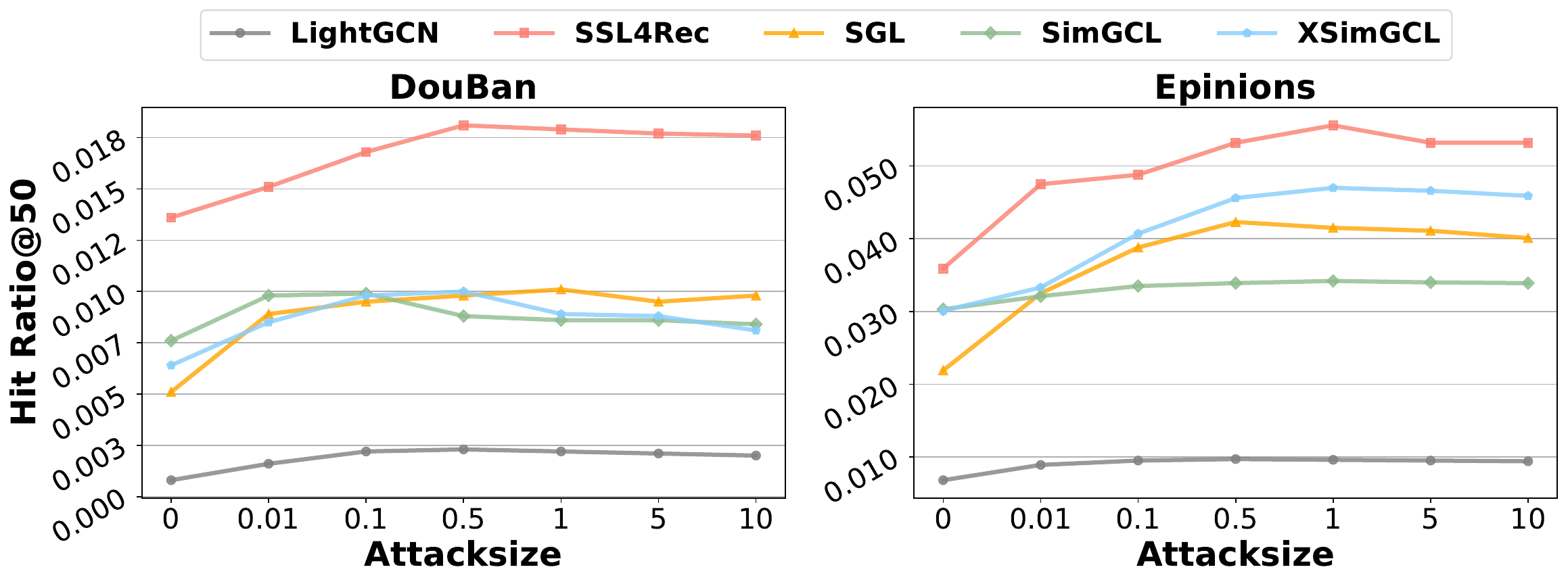}
  \caption{Parameter sensitivity with regard to $\alpha$ on DouBan and Epinions.}
  \vspace{-1pt}
\label{parameters}
\end{figure}

\end{document}